\documentclass{aa}  

\usepackage{graphicx}
\usepackage{txfonts}
\usepackage{placeins}

\begin{document}

   \title{Constraining Effective Viscosity in the Intracluster Medium via the Thermal Sunyaev-Zeldovich Effect}

   \subtitle{Predictions from the SLOW Constrained Coma Cluster Simulations}

   \author{F. Groth\inst{1}
        \and Y. Li\inst{2}
        \and J. Golec\inst{2}
        \and B. A. Seidel\inst{1}
        \and J. G. Sorce\inst{3,4}
        \and T. Marin-Gilabert\inst{1,5}
        \and K. Dolag\inst{1,6}
        }

    \institute{Universit{\"a}ts-Sternwarte, Fakult{\"a}t für Physik, Ludwig-Maximilians-Universit{\"a}t M{\"u}nchen, Scheinerstr. 1, 81679 M{\"u}nchen, Germany\\
    \email{fgroth@usm.lmu.de}
    \and Department of Astronomy, University of Massachusetts, Amherst, MA 01003, USA
    \and Univ. Lille, CNRS, Centrale Lille, UMR 9189 CRIStAL, F-59000 Lille, France
    \and Universit\'e Paris-Saclay, CNRS, Institut d'Astrophysique Spatiale, 91405, Orsay, France
    \and Center for Astrophysics | Harvard \& Smithsonian, 60 Garden St. Cambridge, MA 02138, USA
    \and Max-Planck-Institut f{\"u}r Astrophysik, Karl-Schwarzschild-Stra{\ss}e 1, 85741 Garching, Germany}

   \date{Received September 30, 20XX}
 
  \abstract
   {}
   {We study the effect of viscosity on the Sunyaev-Zeldovich signal in a simulated constrained Coma cluster analog. We aim to provide alternative constraints on the amount of viscosity in the ICM.} 
   {We use the Coma cluster realization with different levels of viscosity from the LOWER DECKS zoom-ins of the SLOW constrained simulations. We generate mock thermal and kinetic Sunyaev-Zeldovich maps and and analyze their statistics. We compare them to Planck observations.}
   {Viscosity shows a consistent trend in thermal SZ (tSZ) profiles, increasing the signal in the center and suppressing it in the outskirts. Viscosity also has a strong effect on the tSZ power spectrum, elevating its amplitude on all scales. Comparisons with Planck observations suggest that the effective ICM viscosity is below $5\%$ of the Spitzer value. Unsharp masking reveals an effect on small scales, which are, however, not yet detectable with current observational data.}
   {The thermal Sunyaev-Zeldovich effect shows clear and consistent trends that allow us to probe the effective viscosity of the ICM. Our analysis suggests suppressed ICM viscosity below $5\%$ of the Spitzer value, consistent with previous X-ray analysis. Our results validate the strength of the SZ effect as an independent method to constrain ICM viscosity.}


   \maketitle

\section{Introduction}

The intracluster medium (ICM) of galaxy clusters is a highly dynamic environment, constantly stirred by accretion and mergers, connected to small scales via a turbulent cascade \citep{Kravtsov&Borgani2012,Angelinelli+2020,Mohapatra+2021,Sayers+2021}. The dynamical properties of the ICM thus depend on the dynamical state of the cluster \citep{Lau+2009, Groth+2025a}. 

Idealized models based on Coulomb collisionality predict that viscosity would affect the turbulent cascade and alter gas dynamics such as the development of mixing instabilities. No evidence for steepening of the turbulent cascade was found in galaxy clusters, however \citep{Zhuravleva+2019}. 
Direct plasma simulations reveal possible reasons for this discrepancy \citep{Donnert+2018}.
An enhancement of collisionality and subsequent suppression of viscosity could be caused by magnetic fields \citep{Squire+2019} or plasma micro-instabilities \citep{Kunz+2014} such as the ``firehose'' or ``mirror'' instability.
The amount of viscosity in the ICM is thus still under debate.

Traditionally, X-ray observations have been used to constrain the amount of viscosity in the ICM.
X-ray surface brightness fluctuations can be linked to turbulent density fluctuations. \citet{Zhuravleva+2019} find strongly suppressed viscosity compared to the assumption of Coulomb collisions \citep[][``Spitzer viscosity'']{Spitzer1962}. Nevertheless, the interpretation turns out to be more complex, as the viscosity scales with temperature, and only isothermal clusters show clear trends \citep{Marin-Gilabert+2024}.
Alternative constraints on the amount of viscosity come from mixing and cold fronts \citep{ZuHone+2014} and the tails of Jellyfish galaxies \citep{Li+2023b,Ignesti+2024a}, also indicating strongly suppressed viscosity compared to the Spitzer value.

More recently, as a continuation of Hitomi \citep{HitomiCollaboration+2016,HitomiCollaboration+2018}, the XRISM satellite \citep{XRISMScienceTeam2022} also observes velocity dispersions, probing the small-scale ICM velocity structure, directly linked to turbulence. Simulations tend to overestimate the velocity dispersion \citep{XRISMCollaboration+2025d}, which could be attributed to viscosity, but also projection effects \citep{Vazza&Brunetti2025,Lebeau+2026} and selection effects \citep{Groth+2026}, as well as treatment of AGN feedback \citep{XrismCollaboration+2025j}. Additional constraints on the ICM viscosity come from the velocity structure function (VSF), which appears to be too steep compared to ideal Kolmogorov turbulence \citep{XRISMCollaboration+2025d} but is also highly sensitive to recent mergers.
How important viscosity is in shaping ICM dynamics thus remains an open question.

As argued by \citet{Dolag+2005a}, alternative and potentially cleaner constraints on turbulence and the underlying viscosity can be derived from Sunyaev-Zel'dovich (SZ) observations. Turbulence leads to deviations from local thermal pressure equilibrium, thus visible in SZ maps. While this work focused on the artificial viscosity in simulation codes, one can use the same effect to constrain physical viscosity.

In addition, most previous constraints on the amount of viscosity are based on statistical comparisons with observations. With recent improvements in constrained simulations, direct one-to-one comparisons between observations and simulated counterparts are possible.
Constrained initial conditions can be constructed based on different approaches. For Coma, well-matched counterparts exist, constrained using galaxy peculiar velocities \citep[e.g.,][]{Sorce2018,Sorce+2021}, or galaxy densities \citep[e.g.,][]{Steinwandel+2026}.

In this work, we combine two approaches to achieve alternative constraints on the amount of viscosity. We use constrained simulations of the Coma cluster with different amounts of physical viscosity, which allow for a cluster-to-cluster comparison in contrast to the exclusively statistical comparisons carried out in the past.
Based on these simulations, we explore the SZ signatures of viscosity via mock observations, comparing to Planck observations.

In Sec.~\ref{sec:simulations} we describe the initial conditions and the code used for the simulations. We analyze the simulations in Sec.~\ref{sec:mock-observations} based on mock SZ observations. Finally, we conclude and discuss our findings in Sec.~\ref{sec:conclusions}.

\begin{figure*}[t!]
    \centering
    \makebox[\linewidth][l]{\includegraphics[width=\linewidth]{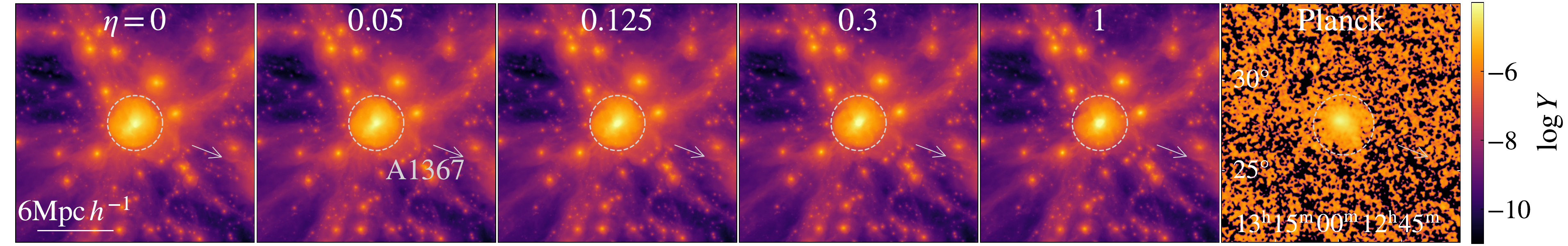}}
    \makebox[\linewidth][l]{\includegraphics[width=0.855\linewidth]{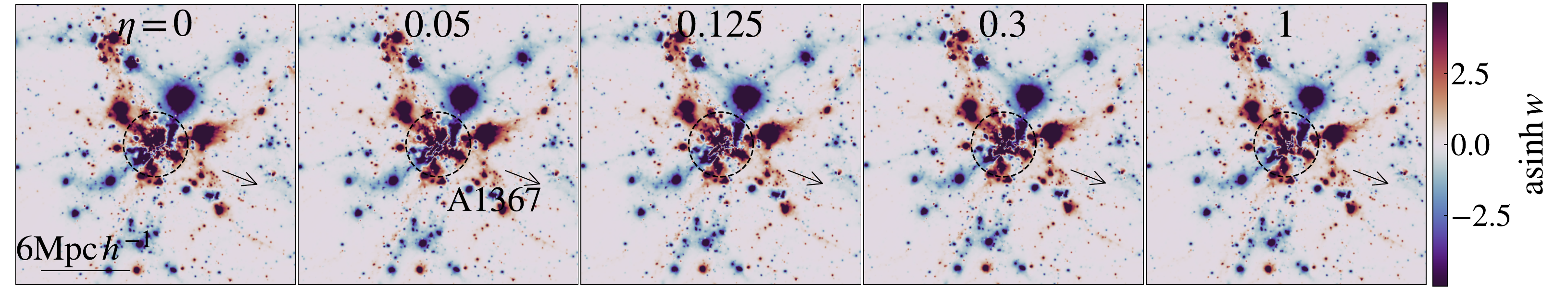}}
    \caption{Top: Mock-observation tSZ $Y$ maps with different amounts of viscosity compared to \citet{PlanckCollaboration+2013,Churazov+2021} observations of the Coma cluster. Bottom: Mock kSZ $w$ maps for the same simulations. The circle indicates $R_{\rm vir}$, based on \textsc{subfind} \citep{Springel+2001,Dolag+2009} for simulated clusters, and the observed value by \citet{Malavasi+2020} for the \citet{PlanckCollaboration+2013} map.}
    \label{fig:maps}
\end{figure*}

\section{The Simulations} \label{sec:simulations}

\subsection{The SLOW Initial Conditions}
The simulations are based on a Coma cluster analog from the LOWER DECKS constrained zoom-in simulations \citep{Seidel+2024,Seidel+2026}, which are extracted from the SLOW cosmological box \citep{Dolag+2023}. The SLOW simulation itself is a realization of the CLONES simulation set \citep{Sorce2018,Sorce+2021}, constructed using peculiar velocities of galaxies to constrain the density field.

These simulations have been shown to very well reproduce the structure of local galaxy clusters \citep{Hernandez-Martinez+2024}, including radial profiles and large-scale environment. In addition, \citet{Groth+2026} showed that the simulations can reproduce the velocity dispersion observed by XRISM, otherwise challenging to model correctly. Thus, they are an ideal starting point to study the effect of viscosity on multi-wavelength observables. As Coma is a very close galaxy cluster, we improve the comparison by using the observer position that best reproduces the position of the real Coma cluster ($[246.980, 245.478, 254.286]$\,Mpc\,$h^{-1}$) instead of the box center.

Zoom-in regions are constructed based on the bound region in the future \citep{Seidel+2024,Seidel+2026}, ensuring no perturbing massive boundary particles are present within $6R_{\rm vir}$. Further, by creating a smooth boundary avoiding edges, this makes the region very stable, even for Godunov methods such as Meshless Finite Mass (MFM).

In this study, we use the Coma cluster analog, simulated at a mass resolution of $M_{\rm gas}=7.2\cdot10^{6}M_{\odot}$ and $M_{\rm DM}=3.9\cdot10^{7}M_{\odot}$ in the high-resolution zoomed region, corresponding to a resolution of the original cosmological box of $2\times6144^3$ particles. Simulations assume a \citet{PlanckCollaboration+2014} background cosmology with $H_0=67.77\,$km\,s$^{-1}$\,Mpc$^{-1}$, $\Omega_{\rm m}=0.307115$, $\Omega_\Lambda=0.692885$, and $\Omega_{\rm b}=0.0480217$.

\subsection{The OpenGadget3 simulation code}
Simulations have been carried out with the \textsc{OpenGadget3} simulation code \citep[][OpenGadget3 Collaboration in prep.]{Groth+2023b}, using the Meshless Finite Mass (MFM) hydro-solver.
This solver has the advantage of very low numerical viscosity, even compared to modern SPH, as no artificial viscosity is required.
MFM evolves the turbulent cascade more accurately, in particular for highly subsonic turbulence as present in the ICM \citep{Groth+2025a}.
Thus, it is the ideal foundation to study the imprint of physical viscosity on the ICM structure, even at the very low amounts expected from previous constraints.

Densities and hydrodynamical forces are evaluated using a cubic spline kernel \citep{Monaghan&Lattanzio1985}.
The physical viscosity implementation is described by \citet{Sijacki&Springel2006,Marin-Gilabert+2022a}.
We vary the amount of viscosity from no physical viscosity to $5\%$, $12.5\%$, $30\%$, and full \citet{Spitzer1962} viscosity $\eta_{\rm Spitzer}$.
To focus on the effect of viscosity, we restrict this analysis to simulations that do not include additional physics, such as cooling or AGN feedback. In addition, we expect these not to be important in the Coma cluster, which shows no recent AGN activity or cooling flows.

\section{Multi-Wavelength Mock-Observations} \label{sec:mock-observations}

In a first step, we create mock-observations using the smac map-making tool \citep{Dolag+2005} for thermal SZ (tSZ) and kinetic SZ (kSZ).
The tSZ effect traces the electron pressure in the ICM, and the kSZ effect traces the line-of-sight bulk ICM velocity
We ensure that mock SZ-maps have the same format, size, and projection as observed maps by following the same pipeline:
We first produce HEALPix maps with $N_{\rm side}=2048$, as for the \citet{PlanckCollaboration+2013} public data. We then extract a gnomonic projection of a $13.3^\circ\times13.3^\circ$ region, the same extent as the map used by \citet{Churazov+2021}. Overall, this minimizes systematic differences in the map construction, allowing for very close one-to-one comparisons.

As a comparison, we show and analyze the \citet{PlanckCollaboration+2013} SZ-map used by \citet{Churazov+2021}. For creating the radial profile directly from this map, we added an offset of $y_{\rm off} = -6.3\cdot 10^{-7}$ as suggested by \citet{PlanckCollaboration+2013}. All values below $Y<y_{\rm off}$ should be treated as unresolved and dominated by background noise. The north direction and corresponding rotation of the simulated maps are determined using the nearest three clusters as described by \citet{Groth+2026}. The correction for the Coma cluster is, however, only $0.04\pi$, indicating the rotation of the simulation is already a good match. 
All maps are shown in J2000 coordinates.

The tSZ $Y$ maps and the kSZ $w$ maps for the simulations with different amounts of viscosity, and, if available, the observed maps, are shown in Fig.~\ref{fig:maps}. The filaments around the simulated Coma cluster analog match the directions of the north and west filaments described by \citet{Malavasi+2020,Malavasi+2023,HyeongHan+2024}. Our simulations even produce an A1367 analog in the east direction \citep{Hernandez-Martinez+2024,Seidel+2024}, indicated in the map by an arrow.
The filament in the south/south-east direction is also present in the simulations, although less strongly pronounced, consistent with the less strong weak lensing signal found by \citet{HyeongHan+2024}. This emphasizes the good match of the cluster, including its galactic environment, important for matching its dynamical state.

On large scales, no clear impact of the amount of viscosity is visible on the maps. A closer investigation of the center indicates that the merger toward higher viscosity is more advanced. We clearly see that the viscosity leads to faster sinking of the two sub-halos associated with the two BCGs towards the cluster center in our simulation, indicating small timing differences of the merger between the different viscosity levels.

\begin{figure*}
    \centering
    \includegraphics[width=0.49\linewidth]{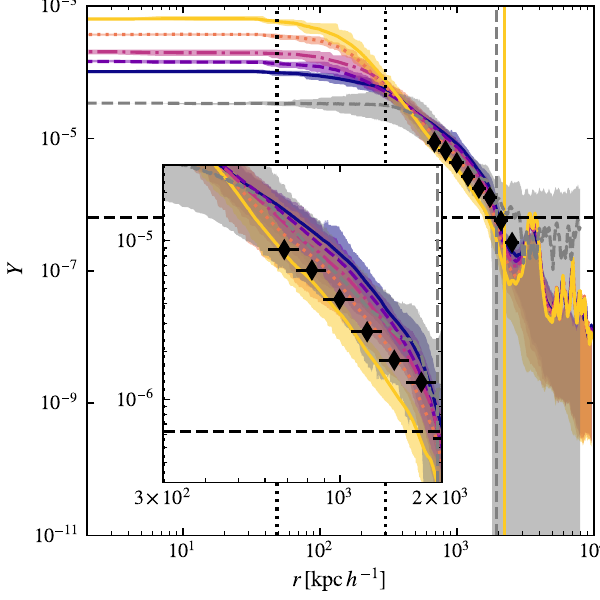}
    \includegraphics[width=0.49\linewidth]{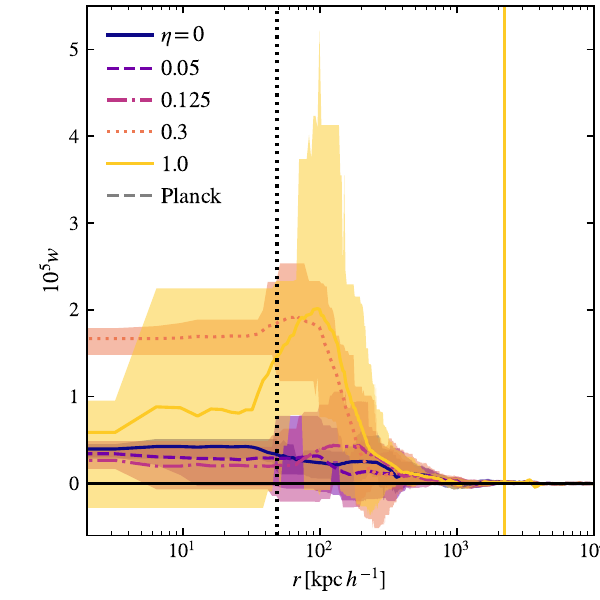}
    \caption{Radial tSZ $Y$ (left) and kSZ $w$ (right) profiles. Different colors denote simulations with different amounts of physical viscosity in units of the Spitzer viscosity; the shaded area indicates the $1\sigma$ uncertainty range. Black diamonds show published profiles by \citet{PlanckCollaboration+2013}. The horizontal line indicates the background level $-y_{\rm off}$, the vertical lines (from left to right): the HEALPix pixel scale, the Planck resolution limit, and the virial radius \citep[observed value by][]{Malavasi+2020}.}
    \label{fig:radial_sz}
\end{figure*}

To compare the data more quantitatively, we extract radial profiles for the SZ maps, shown in Fig.~\ref{fig:radial_sz}. To convert the observed map to physical units, we use the observed redshift $z=0.023997$ \citep{Dhawan+2018}, translating into an angular diameter distance $d\approx103.1$\,Mpc assuming \citet{PlanckCollaboration+2014} cosmology.

The tSZ profile is well measured within $R_{\rm vir}$, whereas further out the observed profile becomes dominated by the background.
Within the virial radius, the tSZ profile yields a clear and consistent trend with viscosity: With increasing amounts of viscosity, central $Y$ values are increased, while in the outskirts, $Y$ values are reduced.
This implies that the gas pressure profile is more centralized in the case of high viscosity.
The increase in pressure arises solely from an increase in density by approximately one order of magnitude between the non-viscous case and the fully viscous case\footnote{Note that this would be even more prominent in X-ray emission (Marin-Gilabert et al. in prep.).}, while the temperature profile remains almost unchanged. This effect has already been shown for numerical viscosity by \citet{Dolag+2005a} as a result of an increased velocity dispersion leading to a turbulent pressure support. Further, turbulent mixing allows for a higher density contrast, and as a consequence for a higher pressure contrast.
The impact on the density makes the effect of viscosity clearly distinct from cooling or feedback, which act primarily on the gas temperature.

In addition, the viscosity has a distinct signature in the entropy profile, as explained by \citet{Dolag+2005a,Vazza+2011} for numerical viscosity.
Shocked high-entropy material from the outskirts of infalling structures can be mixed into the core if viscosity is lower, increasing the central entropy.

Overall, the tSZ profile of our simulated Coma analog is more concentrated than observed by \citet{PlanckCollaboration+2013}.
Nevertheless, the deviation occurs only within the central $r\lesssim 300$\,kpc\,$h^{-1}$, which corresponds to the Planck resolution limit and covers the region where galaxy formation physics would change the hydrodynamical properties in any case. 
In the outskirts beyond $r\gtrsim 2000$\,kpc\,$h^{-1}$, the background starts to dominate. The intermediate range, however, is not affected by either of the effects.
In this regime, higher levels of viscosity lead to lower tSZ signals. Planck observations agree best with $\eta_{\rm Spitzer}>\eta>0.3\eta_{\rm Spitzer}$. The differences are, however, smaller than the scatter.
Further uncertainties arise due to possible deviations in the cluster mass in the simulation, as well as uncertainties in the cluster distance, which both could lead to a rescaling of the profile. Indeed, the difference between the virial radius estimations of Planck observations vs our constrained simulations is similar to the difference between the radial profiles for different levels of viscosity. Considering a possible rescaling of the profiles, the constraints could thus also shift towards even more suppressed viscosity compared to the full Spitzer value.

Overall, the limited resolution and sensitivity of \citet{PlanckCollaboration+2013} observed maps do not allow for deriving any strong constraints from the radial tSZ profiles.
Higher-quality data would be required to derive tighter constraints.
In addition, the effect of viscosity on the tSZ profile is expected to be degenerate with cooling and feedback, not yet included in the simulations.
While cooling could lead to more centralized profiles, AGN feedback could push gas further out, increasing the pressure and thus the tSZ signal in the outskirts.

For the kSZ $w$ signal, no observational data of sufficient resolution are available, as it is much weaker than the tSZ $Y$.
The kSZ effect leads to, on average, zero values, as the mean absolute velocity of the cluster is zero. Deviations are visible in the center within $\lesssim300\,$kpc\,$h^{-1}$, associated with substructure movement, in particular the two BCGs. Nevertheless, no clear trend with viscosity is visible, neither for the absolute value nor for the scatter.

\begin{figure}
    \centering
    \includegraphics[width=0.49\linewidth]{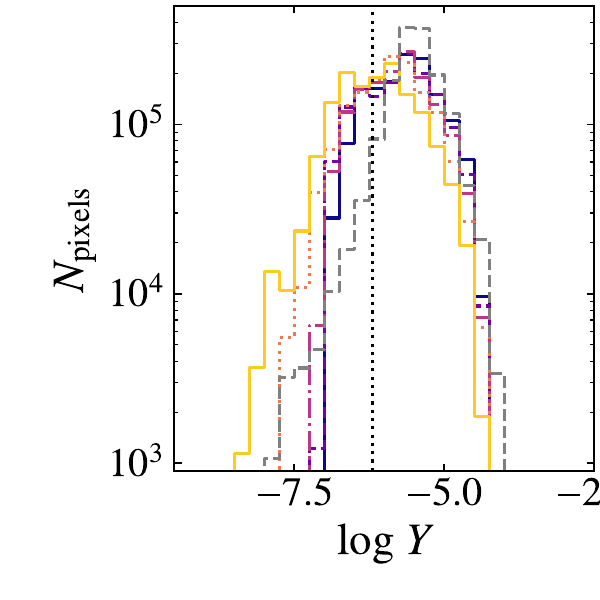}
    \includegraphics[width=0.49\linewidth]{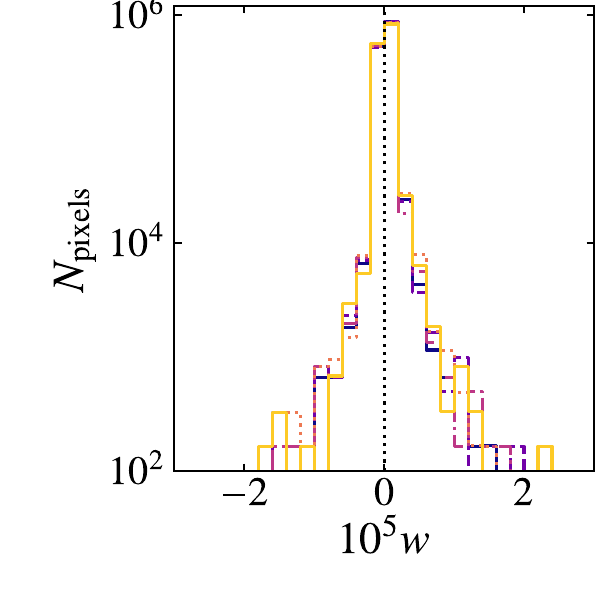}
    \caption{Histograms of the tSZ (left) / kSZ (right) values from the (mock) SZ maps within $0.2R_{\rm vir}<r<R_{\rm vir}$. The vertical line for the tSZ histogram denotes the background level $-y_{\rm off}$, the one for kSZ the zero-value. Same colors as in Fig.~\ref{fig:radial_sz}.}
    \label{fig:sz_hist}
\end{figure}

Differences between different viscosity values can also be characterized via the distribution of $Y$ and $w$ values, shown in Fig.~\ref{fig:sz_hist}.
We focus on the central halo, excluding adjacent substructures, restricting the analysis to $r<R_{\rm vir}$.
For the tSZ $Y$ values, we also exclude the central, resolution-limited range $r<0.2R_{\rm vir}$.
Similarly, for the kSZ $w$ values, we exclude the central signal that is dominated by the BCGs $r<0.2R_{\rm vir}<r$.

In the radial range considered, excluding the central tSZ peak, higher viscosity leads to a more extended tail towards small tSZ $Y$ values, which are, however, below the Planck background limit. Towards higher, resolved values, differences are less significant. Increased viscosity suppresses the higher $Y$ values. The Planck observations in this case even hint towards strongly suppressed viscosity, but higher resolution and sensitivity observations would be required to confirm these trends.

For the kSZ $w$ distribution, all simulations peak at zero, independent of the viscosity, as the ICM on average is at rest. 
The tails of the distribution are broadened for higher levels of viscosity.
While viscosity is expected to reduce turbulence, the kSZ effect is dominated by line-of-sight bulk motions, which are affected by viscosity more indirectly. Substructures mix less and are destroyed later, leading to more coherent gas motions and higher kSZ signals.
This also leads to a strong dependence of the effect on the choice of the region. If excluding more of the center, restricting to $r>0.3R_{\rm vir}$ the difference becomes much less pronounced, and for $r>0.5R_{\rm vir}$ the difference between different levels of viscosity vanishes.
Overall, the necessary resolution to observe these differences cannot be reached in current observations. While constraints might be derived from maximum $w$ values reached, these are also strongly influenced by the timing of the merger and the configuration of the two BCGs, making constraints non-trivial.

\begin{figure*}
    \centering
    \includegraphics[width=0.49\linewidth]{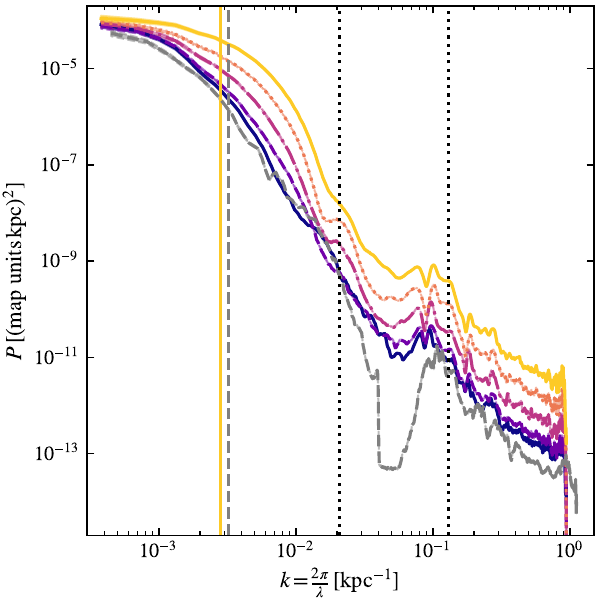}
    \includegraphics[width=0.49\linewidth]{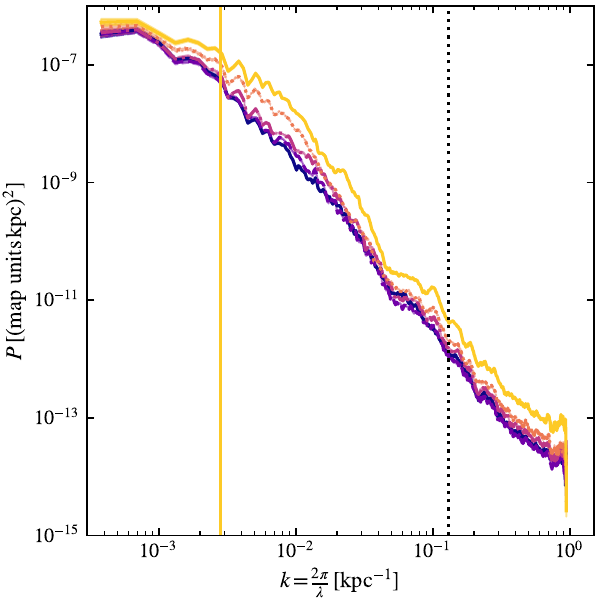}
    \caption{Power spectrum calculated from the (mock) SZ images of the tSZ (left) / kSZ (right) data. Same colors as in Fig.~\ref{fig:radial_sz}. The vertical lines indicate (from left to right) the virial radius, the Planck resolution limit, and the HEALPix pixel scale.}
    \label{fig:powerspectrum}
\end{figure*}

To analyze the structure of the SZ signal, we calculate the 1D power spectrum shown in Fig.~\ref{fig:powerspectrum}. 
We assume Cartesian geometry using the FFTW Julia library, averaging within spherical bins. The gnomonic projection introduces slight deformations compared to the Cartesian layout assumed, but these deformations are limited to a few percent for the Coma field of view. 
As the effect is present both for simulations and observations, it will furthermore not impact the comparison and possible constraints. We add zero-padding and use a Hanning window function to avoid boundary artifacts.

Overall, a good agreement between the simulated and observed power spectra is achieved.
The Planck resolution of $\approx 300$\,kpc corresponds to $k\approx0.02$\,kpc$^{-1}$, for smaller scales (larger $k$), the observed spectrum is dominated by the Planck beam window function, causing an initial suppression and noise leading to amplification for even smaller scales.

Large scales beyond the central halo size are mostly unaffected by viscosity.
The main constraints can be derived from the one-halo term between $3\cdot10^{-3}\,{\rm kpc}^{-1}<k<2\cdot10^{-2}\,{\rm kpc}^{-1}$.

Higher viscosity leads to a higher amplitude in pressure fluctuations, while the slope of the power spectrum remains unaffected. In addition, the breaking scale where the transition from bulk motions to turbulence occurs, shifts towards smaller scales with increased viscosity. As mixing is suppressed, coherent structures are destroyed later, and thus more power remains on larger scales.

The power spectrum computed on the observed Planck tSZ map agrees best with strongly suppressed viscosity $\eta\lesssim0.05\eta_{\rm Spitzer}$. Differences between the power spectra are much larger than between radial profiles, making them a promising analysis to constrain the ICM viscosity.
Nevertheless, as this signal is dominated by the central halo, additional physical processes, in particular feedback, can potentially alter the central power spectrum. We expect differences on larger scales to be more stable against the impact of cooling and feedback, however.

Also the kSZ power spectrum shows a similar trend with viscosity, even though less pronounced. This emphasizes that the tSZ signal, which is both better observable and shows stronger trends, yields better constraints on the amount of viscosity.

\begin{figure*}
    \centering
    \makebox[\linewidth][l]{\includegraphics[width=\linewidth]{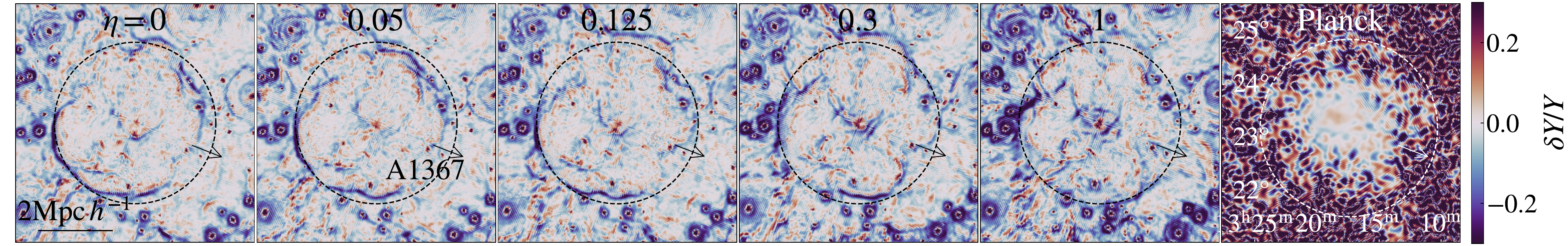}}
    \makebox[\linewidth][l]{\includegraphics[width=0.855\linewidth]{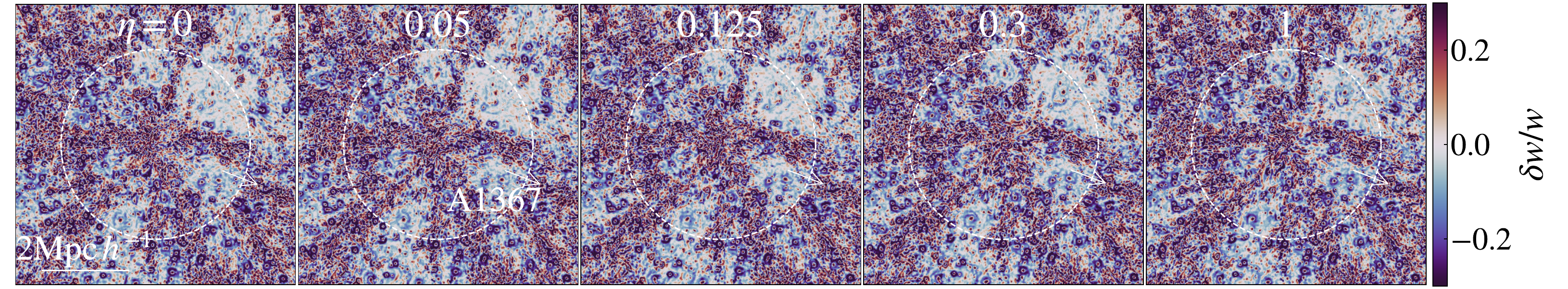}}
    \caption{Unsharp masking maps of tSZ $Y$ (top) and kSZ $w$ (bottom) at a scale $\sigma=100$\,kpc with increasing amount of viscosity from left to right, zoomed into the center compared to Fig.~\ref{fig:maps}. Bottom:.
    A clear impact of the viscosity on the small-scale structure is visible for the tSZ map e.g. for the accretion shocks at $R_{\rm vir}$, while for kSZ no differences are visible.}
    \label{fig:unsharp_masking}
\end{figure*}

An alternative to studying the large-scale impact of viscosity on the SZ maps is to focus on small scales, removing the large-scale contribution with unsharp masking \citep{Dolag+2005a}.
The Gaussian-smoothed image at scale $\sigma$ is substracted from the original image, and the result divided by the smoothed map, This reveals relative differences on scales smaller than the smoothing scale.
We use $\sigma=100\,$kpc, which we found yields the clearest differences between the maps, shown in Fig.~\ref{fig:unsharp_masking}.
As Planck's resolution limit is above $100$\,kpc, the resulting map is dominated by noise. We thus exclude this map from the remaining analysis.

In the simulated tSZ maps, pressure jumps (shocks) are visible as blue stripes. 
In the center, the configuration of the two BCGs (positive $\delta Y/Y$) and the corresponding shock signatures (negative $\delta Y/Y$) change due to timing differences.
The accretion shock further out at $\approx R_{\rm vir}$ shows more systematic trends with viscosity:
While the position of the shocks does not depend on the amount of viscosity, the shock appears thinner, and fluctuations are present at smaller scales for lower amounts of viscosity, compared to larger, more coherent structures in the high-viscosity case. This is consistent with the increased large-scale power in the power spectrum.
Overall, higher viscosity leads to, on average, higher fluctuations, consistent with \citet{Marin-Gilabert+2024}.

The kSZ map shows no clear trend with viscosity, and the signal appears to be dominated by noise due to small-scale motions. The power spectrum of the map is consistent with white noise.

\begin{figure}
    \centering
    \includegraphics[width=\linewidth]{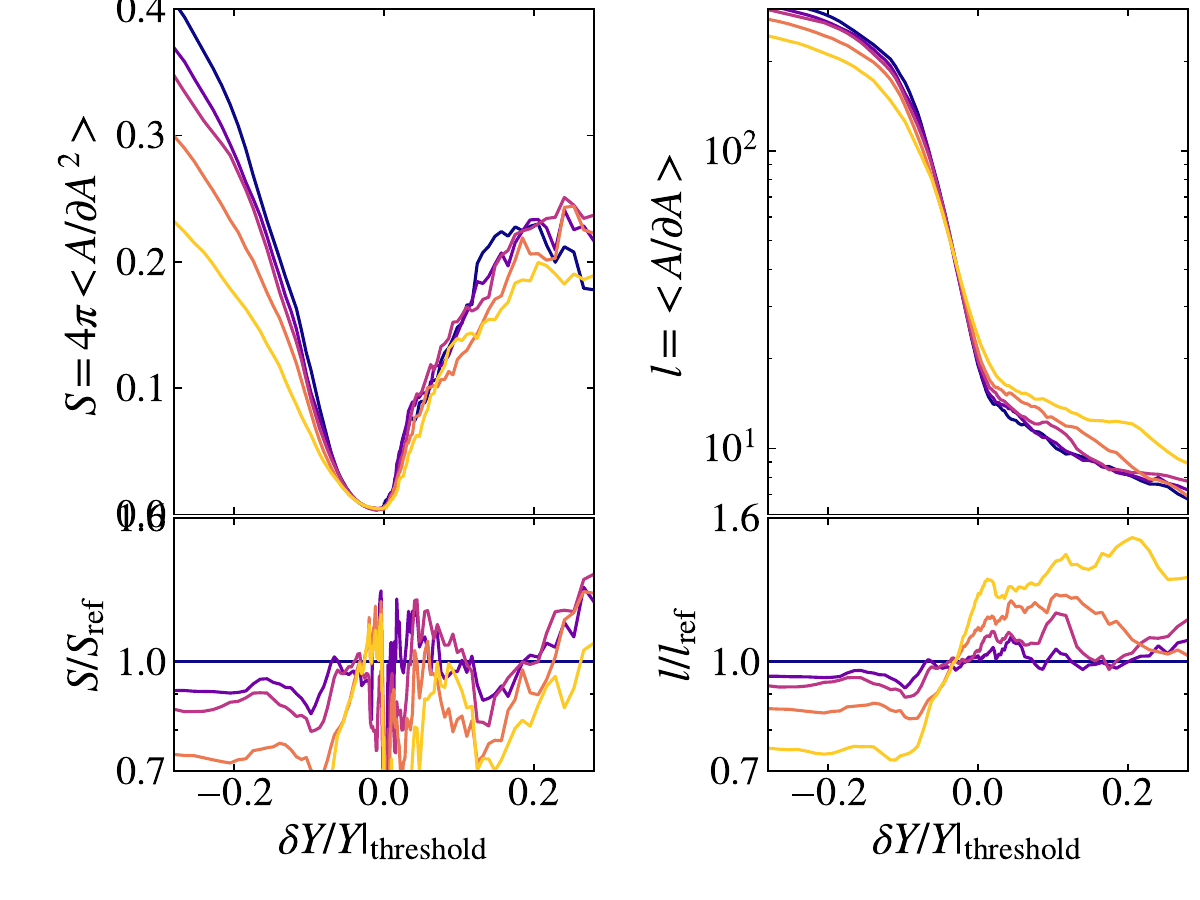}
    \caption{Shape parameter $S$ (left) and coherence length $l$ (right) of the inverted unsharp-masked tSZ maps.}
    \label{fig:morphology_measures}
\end{figure}

Characterizing the difference visible in the shape of structured in the unsharp-masked tSZ map is, however, not trivial.
We find that viscosity produces smaller differences in the power spectrum of the unsharp-masked maps than in the original image.
A clear signature can be found characterizing the shape via Minkowski functionals, in particular the isoperimetric ratio and the coherence length.
In Fig.~\ref{fig:morphology_measures}, we show both quantities calculated for different thresholds.
We degraded the resolution by a factor $16$ to avoid a strong fragmentation of structures. In addition, we inverted the map to focus on the shocks that lead to negative filtered $\delta Y/Y$.

The isoperimetric ratio $S$ characterizes the shape of structures. $S\to 1$ indicates circular structures, while elongated or filamentary structures lead to $S<1$.
In all cases, structures are non-circular ($S\ll1$). Even though structures for the high-viscosity case appear more circular by eye, this is not reflected in the Minkowski-functional-based analysis. This is the result of a complex surface geometry that leads to fragmentation even for reduced resolution and increases the surface estimation for individual connected regions.

The coherence length $l$ describes the typical size of structures, e.g., half the radius for a disk, and half the width for a filament. 
For intermediate to large thresholds $0<\delta Y/Y<0.2$, which capture the shocks, increased viscosity leads to much larger coherence lengths. Structures are on average larger by a factor $1.5$ for $\eta=\eta_{\rm Spitzer}$ compared to $\eta=0$, quantifying the by-eye analysis.

\section{Conclusions} \label{sec:conclusions}

We have simulated a constrained Coma cluster analog, taken from the SLOW/LOWER DECKS initial conditions, and compared mock SZ maps to \citet{PlanckCollaboration+2013} data.

Viscosity has a clear and consistent impact on the tSZ signal. It also shows some effects on the kSZ signal, but the differences for all analysis performed in this work are much smaller and less consistent compared to the tSZ signal. More coherent motions are visible for higher viscosity levels, as structures are destroyed later, but this strongly depends on the region analyzed as well as timing differences. In addition, such small differences in the kSZ signal would be very difficult to measure with current instruments. Possibly, the underlying differences in gas dynamics could be measured indirectly using X-ray (Marin-Gilabert et al. in prep.).

The tSZ signal can be used to derive constraints on the amount of viscosity.
The simulated tSZ profiles show an increased central signal and suppressed signal in the outskirts with increased viscosity. While the current observations from Planck hint towards suppressed viscosity $\eta\lesssim0.3\eta_{\rm Spitzer}$, their constraining power is limited due to Planck's limited resolution. Considering additional uncertainties in the cluster mass and distance hints towards even stronger suppression of the effective ICM viscosity.
The power spectrum shows more observable trends, making it a very promising analysis to put constraints on the ICM viscosity.
For our Coma simulation, the comparison with Planck observations indicated strongly suppressed viscosity $\eta<0.05\eta_{\rm Spitzer}$, consistent with constraints from X-ray measurements \citep{ZuHone+2014,Zhuravleva+2019}.

Nevertheless, some degeneracy of the effect of viscosity with feedback strength is expected. Feedback can push gas further out, increase the pressure, and alter tSZ profiles \citep{Kim+2022,Tanimura+2022}.
Viscosity alters mainly the density profile, while cooling and feedback also affect the temperature profile. Thus, multi-wavelength observations, including tSZ and X-ray, which scale differently with temperature and density, could help to disentangle their effect.
To fully understand the interaction of viscosity with cooling and feedback, more detailed simulations including these processes would be necessary.

Finally, we find that the effect of viscosity on small scale SZ signals can be detected using unsharp-masking. 
Shocks are larger and more coherent for higher levels of viscosity.
As this mainly measures accretion shocks in the outskirts, we do not expect them to change if feedback processes were included.

The $\leq100\,$kpc spatial scales studied via the unsharp-masking correspond to angular scales of $\approx3.4'$, below the $10'$ resolution of Planck. Thus, current observations cannot probe such small scales, resulting in noise-dominated, close-to-zero values for the unsharp-masked map produced based on the data by \citet{PlanckCollaboration+2013}.
Current and forthcoming ground-based millimeter-wave observatories and cameras such as the Simons Observatory \citep{Galitzki+2018}, NIKA2 \citep{Perotto+2024}, and TolTEC \citep{Wilson+2020} are able to reach respective resolutions of roughly $1'$, $20''$, and $10''$.
Observations at those angular scales of the Coma cluster outskirts open up the possibility of constraining the effects of viscosity on shocks. The limited field-of-view of those instruments and the necessity to filter out atmospheric contamination with ground-based observations make it difficult to constrain the Coma cluster in its entirety. Nevertheless, targeted observations of the cluster center and outskirts by those instruments, and leveraging their sensitivities at different angular scales is a promising avenue to constrain the tSZ power spectrum and the structure of shocks at small scales, potentially at the level needed to discriminate between different ICM viscosities.

\begin{acknowledgements}
We thank E. Churazov for providing the observational data used as a reference.

FG, KD, BAS, and TM acknowledge support by the COMPLEX project from the European Research Council (ERC) under the European Union’s Horizon 2020 research and innovation program grant agreement ERC-2019-AdG 882679.
FG and KD acknowledge support by the Deutsche Forschungsgemeinschaft (DFG, German Research Foundation) under Germany’s Excellence Strategy - EXC-2094 - 390783311.
YL acknowledges support from NASA grant 80NSSC22K0668, Chandra X-ray Observatory grant TM3-24005X, NSF grant AST-2510198, and CAREER award AST-2516092.
JG is supported by the National Science Foundation under Award No. 2401781.
KD, JS, and BAS acknowledge support by the grant agreements ANR-21-CE31-0019 / 490702358 from the French Agence Nationale de la Recherche / DFG for the LOCALIZATION project.
This work was supported by the French government under the France 2030 program with the reference ANR-21-IDES-0006. The Métropole Européenne de Lille and the University of Lille are also gratefully acknowledged for the funding and support granted to the WILL-CHAIRES-25-009-UNIVERSITWINS project.
TM acknowledges the support provided by a Smithsonian Scholar Award.
\end{acknowledgements}

\bibliographystyle{aa}
\bibliography{bibliography.bib}

\begin{thebibliography}{51}
\expandafter\ifx\csname natexlab\endcsname\relax\def\natexlab#1{#1}\fi

\bibitem[{Angelinelli {et~al.}(2020)Angelinelli, Vazza, Giocoli, Ettori, Jones, Brunetti, Br{\"u}ggen, \& Eckert}]{Angelinelli+2020}
Angelinelli, M., Vazza, F., Giocoli, C., {et~al.} 2020, Mon. Not. R. Astron. Soc., 495, 864

\bibitem[{Churazov {et~al.}(2021)Churazov, Khabibullin, Lyskova, Sunyaev, \& Bykov}]{Churazov+2021}
Churazov, E., Khabibullin, I., Lyskova, N., Sunyaev, R., \& Bykov, A.~M. 2021, Astron. Astrophys., 651, A41

\bibitem[{Dhawan {et~al.}(2018)Dhawan, Jha, \& Leibundgut}]{Dhawan+2018}
Dhawan, S., Jha, S.~W., \& Leibundgut, B. 2018, Astron. Astrophys., 609, A72

\bibitem[{Dolag {et~al.}(2009)Dolag, Borgani, Murante, \& Springel}]{Dolag+2009}
Dolag, K., Borgani, S., Murante, G., \& Springel, V. 2009, Mon. Not. R. Astron. Soc., 399, 497

\bibitem[{Dolag {et~al.}(2005{\natexlab{a}})Dolag, Hansen, Roncarelli, \& Moscardini}]{Dolag+2005}
Dolag, K., Hansen, F.~K., Roncarelli, M., \& Moscardini, L. 2005{\natexlab{a}}, Mon. Not. R. Astron. Soc., 363, 29

\bibitem[{Dolag {et~al.}(2023)Dolag, Sorce, Pilipenko, {Hern{\'a}ndez-Mart{\'i}nez}, Valentini, Gottl{\"o}ber, Aghanim, \& Khabibullin}]{Dolag+2023}
Dolag, K., Sorce, J.~G., Pilipenko, S., {et~al.} 2023, Astron. Astrophys., 677, A169

\bibitem[{Dolag {et~al.}(2005{\natexlab{b}})Dolag, Vazza, Brunetti, \& Tormen}]{Dolag+2005a}
Dolag, K., Vazza, F., Brunetti, G., \& Tormen, G. 2005{\natexlab{b}}, Mon. Not. R. Astron. Soc., 364, 753

\bibitem[{Donnert {et~al.}(2018)Donnert, Vazza, Br{\"u}ggen, \& ZuHone}]{Donnert+2018}
Donnert, J., Vazza, F., Br{\"u}ggen, M., \& ZuHone, J. 2018, Space Sci. Rev., 214, 122

\bibitem[{Galitzki {et~al.}(2018)Galitzki, Ali, Arnold, Ashton, Austermann, Baccigalupi, Baildon, Barron, Beall, Beckman, Bruno, Bryan, Calisse, Chesmore, Chinone, Choi, Coppi, Crowley, Crowley, Cukierman, Devlin, Dicker, Dober, Duff, Dunkley, Fabbian, Gallardo, Gerbino, {Goeckner-Wald}, Golec, Gudmundsson, Healy, Henderson, Hill, Hilton, Ho, Howe, Hubmayr, Jeong, Keating, Koopman, Kiuchi, Kusaka, Lashner, Lee, Li, Limon, Lungu, Matsuda, Mauskopf, May, McCallum, McMahon, Nati, Niemack, {Orlowski-Scherer}, Parshley, Piccirillo, Sathyanarayana~Rao, Raum, Salatino, Seibert, Sierra, {Silva-Feaver}, Simon, Staggs, Stevens, Suzuki, Teply, Thornton, Tsai, Ullom, Vavagiakis, Vissers, Westbrook, Wollack, Xu, \& Zhu}]{Galitzki+2018}
Galitzki, N., Ali, A., Arnold, K.~S., {et~al.} 2018, in Millimeter, {{Submillimeter}}, and {{Far-Infrared Detectors}} and {{Instrumentation}} for {{Astronomy IX}}, Vol. 10708, eprint: arXiv:1808.04493, 1070804

\bibitem[{Groth {et~al.}(2023)Groth, Steinwandel, Valentini, \& Dolag}]{Groth+2023b}
Groth, F., Steinwandel, U.~P., Valentini, M., \& Dolag, K. 2023, Mon. Not. R. Astron. Soc., 526, 616

\bibitem[{Groth {et~al.}(2026)Groth, Valentini, Seidel, {Vladutescu-Zopp}, Biffi, Dolag, \& Sorce}]{Groth+2026}
Groth, F., Valentini, M., Seidel, B.~A., {et~al.} 2026, Astrophys. J., 1000, 75

\bibitem[{Groth {et~al.}(2025)Groth, Valentini, Steinwandel, {Vall{\'e}s-P{\'e}rez}, \& Dolag}]{Groth+2025a}
Groth, F., Valentini, M., Steinwandel, U.~P., {Vall{\'e}s-P{\'e}rez}, D., \& Dolag, K. 2025, Astron. Astrophys., 693, A263

\bibitem[{{Hern{\'a}ndez-Mart{\'i}nez} {et~al.}(2024){Hern{\'a}ndez-Mart{\'i}nez}, Dolag, Seidel, Sorce, Aghanim, Pilipenko, Gottl{\"o}ber, Lebeau, \& Valentini}]{Hernandez-Martinez+2024}
{Hern{\'a}ndez-Mart{\'i}nez}, E., Dolag, K., Seidel, B., {et~al.} 2024, Astron. Astrophys., 687, A253

\bibitem[{{Hitomi Collaboration} {et~al.}(2016){Hitomi Collaboration}, Aharonian, Akamatsu, Akimoto, Allen, Anabuki, Angelini, Arnaud, Audard, Awaki, Axelsson, Bamba, Bautz, Blandford, Brenneman, Brown, Bulbul, Cackett, Chernyakova, Chiao, Coppi, Costantini, {de Plaa}, {den Herder}, Done, Dotani, Ebisawa, Eckart, Enoto, Ezoe, Fabian, Ferrigno, Foster, Fujimoto, Fukazawa, Furuzawa, Galeazzi, Gallo, Gandhi, Giustini, Goldwurm, Gu, Guainazzi, Haba, Hagino, Hamaguchi, Harrus, Hatsukade, Hayashi, Hayashi, Hayashida, Hiraga, Hornschemeier, Hoshino, Hughes, Iizuka, Inoue, Inoue, Ishibashi, Ishida, Ishikawa, Ishisaki, Itoh, Iyomoto, Kaastra, Kallman, Kamae, Kara, Kataoka, Katsuda, Katsuta, Kawaharada, Kawai, Kelley, Khangulyan, Kilbourne, King, Kitaguchi, Kitamoto, Kitayama, Kohmura, Kokubun, Koyama, Koyama, Kretschmar, Krimm, Kubota, Kunieda, Laurent, Lebrun, Lee, Leutenegger, Limousin, Loewenstein, Long, Lumb, Madejski, Maeda, Maier, Makishima, Markevitch, Matsumoto, Matsushita, McCammon, McNamara, Mehdipour,
  Miller, Miller, Mineshige, Mitsuda, Mitsuishi, Miyazawa, Mizuno, Mori, Mori, Moseley, Mukai, Murakami, Murakami, Mushotzky, Nagino, Nakagawa, Nakajima, Nakamori, Nakano, Nakashima, Nakazawa, Nobukawa, Noda, Nomachi, O'Dell, Odaka, Ohashi, Ohno, Okajima, Ota, Ozaki, Paerels, Paltani, Parmar, Petre, Pinto, Pohl, Porter, Pottschmidt, Ramsey, Reynolds, Russell, {Safi-Harb}, Saito, Sakai, Sameshima, Sato, Sato, Sato, Sawada, Schartel, Serlemitsos, Seta, Shidatsu, Simionescu, Smith, Soong, Stawarz, Sugawara, Sugita, Szymkowiak, Tajima, Takahashi, Takahashi, Takeda, Takei, Tamagawa, Tamura, Tamura, Tanaka, Tanaka, Tanaka, Tashiro, Tawara, Terada, Terashima, Tombesi, Tomida, Tsuboi, Tsujimoto, Tsunemi, Tsuru, Uchida, Uchiyama, Uchiyama, Ueda, Ueda, Ueno, Uno, Urry, Ursino, {de Vries}, Watanabe, Werner, Wik, Wilkins, Williams, Yamada, Yamaguchi, Yamaoka, Yamasaki, Yamauchi, Yamauchi, Yaqoob, Yatsu, Yonetoku, Yoshida, Yuasa, Zhuravleva, \& Zoghbi}]{HitomiCollaboration+2016}
{Hitomi Collaboration}, Aharonian, F., Akamatsu, H., {et~al.} 2016, Nature, 535, 117

\bibitem[{{Hitomi Collaboration} {et~al.}(2018){Hitomi Collaboration}, Aharonian, Akamatsu, Akimoto, Allen, Angelini, Audard, Awaki, Axelsson, Bamba, Bautz, Blandford, Brenneman, Brown, Bulbul, Cackett, Canning, Chernyakova, Chiao, Coppi, Costantini, {de Plaa}, {de Vries}, {den Herder}, Done, Dotani, Ebisawa, Eckart, Enoto, Ezoe, Fabian, Ferrigno, Foster, Fujimoto, Fukazawa, Furuzawa, Galeazzi, Gallo, Gandhi, Giustini, Goldwurm, Gu, Guainazzi, Haba, Hagino, Hamaguchi, Harrus, Hatsukade, Hayashi, Hayashi, Hayashi, Hayashida, Hiraga, Hornschemeier, Hoshino, Hughes, Ichinohe, Iizuka, Inoue, Inoue, Inoue, Ishida, Ishikawa, Ishisaki, Iwai, Kaastra, Kallman, Kamae, Kataoka, Katsuda, Kawai, Kelley, Kilbourne, Kitaguchi, Kitamoto, Kitayama, Kohmura, Kokubun, Koyama, Koyama, Kretschmar, Krimm, Kubota, Kunieda, Laurent, Lee, Leutenegger, Limousin, Loewenstein, Long, Lumb, Madejski, Maeda, Maier, Makishima, Markevitch, Matsumoto, Matsushita, McCammon, McNamara, Mehdipour, Miller, Miller, Mineshige, Mitsuda, Mitsuishi,
  Miyazawa, Mizuno, Mori, Mori, Mukai, Murakami, Mushotzky, Nakagawa, Nakajima, Nakamori, Nakashima, Nakazawa, Nobukawa, Nobukawa, Noda, Odaka, Ohashi, Ohno, Okajima, Ota, Ozaki, Paerels, Paltani, Petre, Pinto, Porter, Pottschmidt, Reynolds, {Safi-Harb}, Saito, Sakai, Sasaki, Sato, Sato, Sato, Sawada, Schartel, Serlemtsos, Seta, Shidatsu, Simionescu, Smith, Soong, Stawarz, Sugawara, Sugita, Szymkowiak, Tajima, Takahashi, Takahashi, Takeda, Takei, Tamagawa, Tamura, Tanaka, Tanaka, Tanaka, Tanaka, Tashiro, Tawara, Terada, Terashima, Tombesi, Tomida, Tsuboi, Tsujimoto, Tsunemi, Tsuru, Uchida, Uchiyama, Uchiyama, Ueda, Ueda, Uno, Urry, Ursino, Wang, Watanabe, Werner, Wilkins, Williams, Yamada, Yamaguchi, Yamaoka, Yamasaki, Yamauchi, Yamauchi, Yaqoob, Yatsu, Yonetoku, Zhuravleva, \& Zoghbi}]{HitomiCollaboration+2018}
{Hitomi Collaboration}, Aharonian, F., Akamatsu, H., {et~al.} 2018, Publ. Astron. Soc. Jpn., 70, 9

\bibitem[{HyeongHan {et~al.}(2024)HyeongHan, Jee, Cha, \& Cho}]{HyeongHan+2024}
HyeongHan, K., Jee, M.~J., Cha, S., \& Cho, H. 2024, Nat. Astron., 8, 377

\bibitem[{Ignesti {et~al.}(2024)Ignesti, Brunetti, Gullieuszik, Akerman, Marasco, Poggianti, Li, Vulcani, Gitti, Moretti, Giunchi, Tomi{\v c}i{\'c}, Bacchini, Paladino, Radovich, \& Wolter}]{Ignesti+2024a}
Ignesti, A., Brunetti, G., Gullieuszik, M., {et~al.} 2024, Astrophys. J., 977, 219

\bibitem[{Kim {et~al.}(2022)Kim, Golwala, Bartlett, Amodeo, Battaglia, Benson, Hill, Hopkins, Hummels, Moser, \& Orr}]{Kim+2022}
Kim, J., Golwala, S., Bartlett, J.~G., {et~al.} 2022, Astrophys. J., 926, 179

\bibitem[{Kravtsov \& Borgani(2012)}]{Kravtsov&Borgani2012}
Kravtsov, A.~V. \& Borgani, S. 2012, Annu. Rev. Astron. Astrophys., 50, 353

\bibitem[{Kunz {et~al.}(2014)Kunz, Schekochihin, \& Stone}]{Kunz+2014}
Kunz, M.~W., Schekochihin, A.~A., \& Stone, J.~M. 2014, Phys. Rev. Lett., 112, 205003

\bibitem[{Lau {et~al.}(2009)Lau, Kravtsov, \& Nagai}]{Lau+2009}
Lau, E.~T., Kravtsov, A.~V., \& Nagai, D. 2009, Astrophys. J., 705, 1129

\bibitem[{Lebeau {et~al.}(2026)Lebeau, Ettori, Sorce, Aghanim, \& Past{\'e}}]{Lebeau+2026}
Lebeau, T., Ettori, S., Sorce, J.~G., Aghanim, N., \& Past{\'e}, J. 2026, Astron. Astrophys., 707, A336

\bibitem[{Li {et~al.}(2023)Li, Luo, Fossati, Sun, \& J{\'a}chym}]{Li+2023b}
Li, Y., Luo, R., Fossati, M., Sun, M., \& J{\'a}chym, P. 2023, Mon. Not. R. Astron. Soc., 521, 4785

\bibitem[{Malavasi {et~al.}(2020)Malavasi, Aghanim, Tanimura, Bonjean, \& Douspis}]{Malavasi+2020}
Malavasi, N., Aghanim, N., Tanimura, H., Bonjean, V., \& Douspis, M. 2020, Astron. Astrophys., 634, A30

\bibitem[{Malavasi {et~al.}(2023)Malavasi, Sorce, Dolag, \& Aghanim}]{Malavasi+2023}
Malavasi, N., Sorce, J.~G., Dolag, K., \& Aghanim, N. 2023, Astron. Astrophys., 675, A76

\bibitem[{{Marin-Gilabert} {et~al.}(2024){Marin-Gilabert}, Steinwandel, Valentini, {Vall{\'e}s-P{\'e}rez}, \& Dolag}]{Marin-Gilabert+2024}
{Marin-Gilabert}, T., Steinwandel, U.~P., Valentini, M., {Vall{\'e}s-P{\'e}rez}, D., \& Dolag, K. 2024, Density {{Fluctuations}} in the {{Intracluster Medium}}: {{An Attempt}} to {{Constrain Viscosity}} with {{Cosmological Simulations}}

\bibitem[{{Marin-Gilabert} {et~al.}(2022){Marin-Gilabert}, Valentini, Steinwandel, \& Dolag}]{Marin-Gilabert+2022a}
{Marin-Gilabert}, T., Valentini, M., Steinwandel, U.~P., \& Dolag, K. 2022, Mon. Not. R. Astron. Soc., 517, 5971

\bibitem[{Mohapatra {et~al.}(2021)Mohapatra, Federrath, \& Sharma}]{Mohapatra+2021}
Mohapatra, R., Federrath, C., \& Sharma, P. 2021, Mon. Not. R. Astron. Soc., 500, 5072

\bibitem[{Monaghan \& Lattanzio(1985)}]{Monaghan&Lattanzio1985}
Monaghan, J.~J. \& Lattanzio, J.~C. 1985, Astron. Astrophys., 149, 135

\bibitem[{Perotto {et~al.}(2024)Perotto, Adam, Ade, Ajeddig, Andr{\'e}, Artis, Aussel, Barrena, Bartalucci, Beelen, Beno{\^i}t, Berta, Bing, Bourrion, Calvo, Catalano, De~Petris, D{\'e}sert, Doyle, Driessen, Ejlali, Ferragamo, Gomez, Goupy, Hanser, Katsioli, K{\'e}ruzor{\'e}, Kramer, Ladjelate, Lagache, Leclercq, Lestrade, {Mac{\'i}as-P{\'e}rez}, Madden, Maury, Mauskopf, Mayet, Monfardini, {Moyer-Anin}, {Mu{\~n}oz-Echeverr{\'i}a}, Paliwal, Pisano, Pointecouteau, Ponthieu, Pratt, Rev{\'e}ret, Rigby, Ritacco, Romero, Roussel, Ruppin, Schuster, Sievers, Tucker, \& Yepes}]{Perotto+2024}
Perotto, L., Adam, R., Ade, P., {et~al.} 2024, in Mm {{Universe}} 2023 - {{Observing}} the {{Universe}} at Mm {{Wavelengths}}, Vol. 293 (eprint: arXiv:2310.04553: EDP), 00040

\bibitem[{{Planck Collaboration} {et~al.}(2014){Planck Collaboration}, Ade, Aghanim, {Armitage-Caplan}, Arnaud, Ashdown, {Atrio-Barandela}, Aumont, Baccigalupi, Banday, Barreiro, Bartlett, Battaner, Benabed, Beno{\^i}t, {Benoit-L{\'e}vy}, Bernard, Bersanelli, Bielewicz, Bobin, Bock, Bonaldi, Bond, Borrill, Bouchet, Bridges, Bucher, Burigana, Butler, Calabrese, Cappellini, Cardoso, Catalano, Challinor, Chamballu, Chary, Chen, Chiang, Chiang, Christensen, Church, Clements, Colombi, Colombo, Couchot, Coulais, Crill, Curto, Cuttaia, Danese, Davies, Davis, {de Bernardis}, {de Rosa}, {de Zotti}, Delabrouille, Delouis, D{\'e}sert, Dickinson, Diego, Dolag, Dole, Donzelli, Dor{\'e}, Douspis, Dunkley, Dupac, Efstathiou, Elsner, En{\ss}lin, Eriksen, Finelli, Forni, Frailis, Fraisse, Franceschi, Gaier, Galeotta, Galli, Ganga, Giard, Giardino, {Giraud-H{\'e}raud}, Gjerl{\o}w, {Gonz{\'a}lez-Nuevo}, G{\'o}rski, Gratton, Gregorio, Gruppuso, Gudmundsson, Haissinski, Hamann, Hansen, Hanson, Harrison, {Henrot-Versill{\'e}},
  {Hern{\'a}ndez-Monteagudo}, Herranz, Hildebrandt, Hivon, Hobson, Holmes, Hornstrup, Hou, Hovest, Huffenberger, Jaffe, Jaffe, Jewell, Jones, Juvela, Keih{\"a}nen, Keskitalo, Kisner, Kneissl, Knoche, Knox, Kunz, {Kurki-Suonio}, Lagache, L{\"a}hteenm{\"a}ki, Lamarre, Lasenby, Lattanzi, Laureijs, Lawrence, Leach, Leahy, Leonardi, {Le{\'o}n-Tavares}, Lesgourgues, Lewis, Liguori, Lilje, {Linden-V{\o}rnle}, {L{\'o}pez-Caniego}, Lubin, {Mac{\'i}as-P{\'e}rez}, Maffei, Maino, Mandolesi, Maris, Marshall, Martin, {Mart{\'i}nez-Gonz{\'a}lez}, Masi, Massardi, Matarrese, Matthai, Mazzotta, Meinhold, Melchiorri, Melin, Mendes, Menegoni, Mennella, Migliaccio, Millea, Mitra, {Miville-Desch{\^e}nes}, Moneti, Montier, Morgante, Mortlock, Moss, Munshi, Murphy, Naselsky, Nati, Natoli, Netterfield, {N{\o}rgaard-Nielsen}, Noviello, Novikov, Novikov, O'Dwyer, Osborne, Oxborrow, Paci, Pagano, Pajot, Paladini, Paoletti, Partridge, Pasian, Patanchon, Pearson, Pearson, Peiris, Perdereau, Perotto, Perrotta, Pettorino, Piacentini, Piat,
  Pierpaoli, Pietrobon, Plaszczynski, Platania, Pointecouteau, Polenta, Ponthieu, Popa, Poutanen, Pratt, Pr{\'e}zeau, Prunet, Puget, Rachen, Reach, Rebolo, Reinecke, Remazeilles, Renault, Ricciardi, Riller, Ristorcelli, Rocha, Rosset, Roudier, {Rowan-Robinson}, {Rubi{\~n}o-Mart{\'i}n}, Rusholme, Sandri, Santos, Savelainen, Savini, Scott, Seiffert, Shellard, Spencer, Starck, Stolyarov, Stompor, Sudiwala, Sunyaev, Sureau, Sutton, {Suur-Uski}, Sygnet, Tauber, Tavagnacco, Terenzi, Toffolatti, Tomasi, Tristram, Tucci, Tuovinen, T{\"u}rler, Umana, Valenziano, Valiviita, Van~Tent, Vielva, Villa, Vittorio, Wade, Wandelt, Wehus, White, White, Wilkinson, Yvon, Zacchei, \& Zonca}]{PlanckCollaboration+2014}
{Planck Collaboration}, Ade, P. A.~R., Aghanim, N., {et~al.} 2014, Astron. Astrophys., 571, A16

\bibitem[{{Planck Collaboration} {et~al.}(2013){Planck Collaboration}, Ade, Aghanim, Arnaud, Ashdown, {Atrio-Barandela}, Aumont, Baccigalupi, Balbi, Banday, Barreiro, Bartlett, Battaner, Benabed, Beno{\^i}t, Bernard, Bersanelli, Bikmaev, B{\"o}hringer, Bonaldi, Bond, Borrill, Bouchet, Bourdin, Brown, Brown, Burenin, Burigana, Cabella, Cardoso, Carvalho, Catalano, Cay{\'o}n, Chiang, Chon, Christensen, Churazov, Clements, Colafrancesco, Colombo, Coulais, Crill, Cuttaia, Da~Silva, Dahle, Danese, Davis, {de Bernardis}, {de Gasperis}, {de Rosa}, {de Zotti}, Delabrouille, D{\'e}mocl{\`e}s, D{\'e}sert, Dickinson, Diego, Dolag, Dole, Donzelli, Dor{\'e}, D{\"o}rl, Douspis, Dupac, En{\ss}lin, Eriksen, Finelli, {Flores-Cacho}, Forni, Frailis, Franceschi, Frommert, Galeotta, Ganga, {G{\'e}nova-Santos}, Giard, Gilfanov, {Gonz{\'a}lez-Nuevo}, G{\'o}rski, Gregorio, Gruppuso, Hansen, Harrison, {Henrot-Versill{\'e}}, {Hern{\'a}ndez-Monteagudo}, Hildebrandt, Hivon, Hobson, Holmes, Hornstrup, Hovest, Huffenberger, Hurier,
  Jaffe, Jagemann, Jones, Juvela, Keih{\"a}nen, Khamitov, Kneissl, Knoche, Knox, Kunz, {Kurki-Suonio}, Lagache, L{\"a}hteenm{\"a}ki, Lamarre, Lasenby, Lawrence, Le~Jeune, Leonardi, Lilje, {Linden-V{\o}rnle}, {L{\'o}pez-Caniego}, Lubin, {Mac{\'i}as-P{\'e}rez}, Maffei, Maino, Mandolesi, Maris, Marleau, {Mart{\'i}nez-Gonz{\'a}lez}, Masi, Massardi, Matarrese, Matthai, Mazzotta, Mei, Melchiorri, Melin, Mendes, Mennella, Mitra, {Miville-Desch{\^e}nes}, Moneti, Montier, Morgante, Munshi, Murphy, Naselsky, Natoli, {N{\o}rgaard-Nielsen}, Noviello, Novikov, Novikov, Osborne, Pajot, Paoletti, Perdereau, Perrotta, Piacentini, Piat, Pierpaoli, Piffaretti, Plaszczynski, Pointecouteau, Polenta, Ponthieu, Popa, Poutanen, Pratt, Prunet, Puget, Rachen, Rebolo, Reinecke, Remazeilles, Renault, Ricciardi, Riller, Ristorcelli, Rocha, Roman, Rosset, Rossetti, {Rubi{\~n}o-Mart{\'i}n}, Rudnick, Rusholme, Sandri, Savini, Schaefer, Scott, Smoot, Stivoli, Sudiwala, Sunyaev, Sutton, {Suur-Uski}, Sygnet, Tauber, Terenzi, Toffolatti,
  Tomasi, Tristram, Tuovinen, T{\"u}rler, Umana, Valenziano, Van~Tent, Varis, \& Vielva}]{PlanckCollaboration+2013}
{Planck Collaboration}, Ade, P. A.~R., Aghanim, N., {et~al.} 2013, Astron. Astrophys., 554, A140

\bibitem[{Sayers {et~al.}(2021)Sayers, Sereno, Ettori, Rasia, Cui, Golwala, Umetsu, \& Yepes}]{Sayers+2021}
Sayers, J., Sereno, M., Ettori, S., {et~al.} 2021, Mon. Not. R. Astron. Soc., 505, 4338

\bibitem[{Seidel {et~al.}(2026)Seidel, Dolag, \& Sorce}]{Seidel+2026}
Seidel, B., Dolag, K., \& Sorce, J.~G. 2026, Cutting with Precision -- {{Leveraging Collapse Volumes}} to Generate the next Generation of Zoom-in Initial Conditions

\bibitem[{Seidel {et~al.}(2024)Seidel, Dolag, Remus, Sorce, {Hern{\'a}ndez-Mart{\'i}nez}, Khabibullin, \& Aghanim}]{Seidel+2024}
Seidel, B.~A., Dolag, K., Remus, R.-S., {et~al.} 2024, {{SLOW IV}}: {{Not}} All That Is {{Close}} Will {{Merge}} in the {{End}}. {{Superclusters}} and Their {{Lagrangian}} Collapse Regions

\bibitem[{Sijacki \& Springel(2006)}]{Sijacki&Springel2006}
Sijacki, D. \& Springel, V. 2006, Mon. Not. R. Astron. Soc., 371, 1025

\bibitem[{Sorce(2018)}]{Sorce2018}
Sorce, J.~G. 2018, Mon. Not. R. Astron. Soc., 478, 5199

\bibitem[{Sorce {et~al.}(2021)Sorce, Dubois, Blaizot, McGee, Yepes, \& Knebe}]{Sorce+2021}
Sorce, J.~G., Dubois, Y., Blaizot, J., {et~al.} 2021, Mon. Not. R. Astron. Soc., 504, 2998

\bibitem[{Spitzer(1962)}]{Spitzer1962}
Spitzer, L. 1962, Physics of {{Fully Ionized Gases}}

\bibitem[{Springel {et~al.}(2001)Springel, White, Tormen, \& Kauffmann}]{Springel+2001}
Springel, V., White, S. D.~M., Tormen, G., \& Kauffmann, G. 2001, Mon. Not. R. Astron. Soc., 328, 726

\bibitem[{Squire {et~al.}(2019)Squire, Schekochihin, Quataert, \& Kunz}]{Squire+2019}
Squire, J., Schekochihin, A.~A., Quataert, E., \& Kunz, M.~W. 2019, J. Plasma Phys., 85, 905850114

\bibitem[{Steinwandel {et~al.}(2026)Steinwandel, McAlpine, Stiskalek, Pakmor, Springel, Churazov, Khabibullin, Jasche, Lavaux, \& Bryan}]{Steinwandel+2026}
Steinwandel, U.~P., McAlpine, S., Stiskalek, R., {et~al.} 2026, Learning the {{Universe}}: {{Constrained}} Simulations of the {{Coma}} Galaxy Cluster -- {{I}}. {{Radial X-ray}} and {{Compton-y}} Signatures

\bibitem[{Tanimura {et~al.}(2022)Tanimura, Hinshaw, McCarthy, Van~Waerbeke, Aghanim, Ma, Mead, Tr{\"o}ster, Hojjati, \& Moraes}]{Tanimura+2022}
Tanimura, H., Hinshaw, G., McCarthy, I.~G., {et~al.} 2022, in Mm {{Universe}} @ {{NIKA2}} - {{Observing}} the Mm {{Universe}} with the {{NIKA2 Camera}}, Vol. 257 (eprint: arXiv:2111.02088: EDP), 00045

\bibitem[{Vazza \& Brunetti(2025)}]{Vazza&Brunetti2025}
Vazza, F. \& Brunetti, G. 2025, On the Interpretation of {{XRISM X-ray}} Measurements of Turbulence in the Intracluster Medium: A Comparison with Cosmological Simulations

\bibitem[{Vazza {et~al.}(2011)Vazza, Dolag, Ryu, Brunetti, Gheller, Kang, \& Pfrommer}]{Vazza+2011}
Vazza, F., Dolag, K., Ryu, D., {et~al.} 2011, Mon. Not. R. Astron. Soc., 418, 960

\bibitem[{Wilson {et~al.}(2020)Wilson, {Abi-Saad}, Ade, Aretxaga, Austermann, Ban, Bardin, Beall, Berthoud, Bryan, Bussan, Castillo, Chavez, Contente, DeNigris, Dober, Eiben, Ferrusca, Fissel, Gao, Golec, Golina, Gomez, Gordon, Gutermuth, Hilton, Hosseini, Hubmayr, Hughes, Kuczarski, Lee, Lunde, Ma, Mani, Mauskopf, McCrackan, McKenney, McMahon, Novak, Pisano, Pope, Ralston, Rodriguez, {S{\'a}nchez-Arg{\"u}elles}, Schloerb, Simon, Sinclair, Souccar, Torres~Campos, Tucker, Ullom, Van~Camp, Van~Lanen, Velazquez, Vissers, Weeks, \& Yun}]{Wilson+2020}
Wilson, G.~W., {Abi-Saad}, S., Ade, P., {et~al.} 2020, in Millimeter, {{Submillimeter}}, and {{Far-Infrared Detectors}} and {{Instrumentation}} for {{Astronomy X}}, Vol. 11453, 1145302

\bibitem[{{XRISM Collaboration} {et~al.}(2025{\natexlab{a}}){XRISM Collaboration}, Audard, Awaki, Ballhausen, Bamba, Behar, {Boissay-Malaquin}, Brenneman, Brown, Corrales, Costantini, Cumbee, Diaz~Trigo, Done, Dotani, Ebisawa, Eckart, Eckert, Eguchi, Enoto, Ezoe, Foster, Fujimoto, Fujita, Fukazawa, Fukushima, Furuzawa, Gallo, Garc{\'i}a, Gu, Guainazzi, Hagino, Hamaguchi, Hatsukade, Hayashi, Hayashi, Hell, {Hodges-Kluck}, Hornschemeier, Ichinohe, Ishi, Ishida, Ishikawa, Ishisaki, Kaastra, Kallman, Kara, Katsuda, Kanemaru, Kelley, Kilbourne, Kitamoto, Kobayashi, Kohmura, Kubota, Leutenegger, Loewenstein, Maeda, Markevitch, Matsumoto, Matsushita, McCammon, McNamara, Mernier, Miller, Miller, Mitsuishi, Mizumoto, Mizuno, Mori, Mukai, Murakami, Mushotzky, Nakajima, Nakazawa, Ness, Nobukawa, Nobukawa, Noda, Odaka, Ogawa, Ogorza{\l}ek, Okajima, Ota, Paltani, Petre, Plucinsky, Porter, Pottschmidt, Sato, Sato, Sawada, Seta, Shidatsu, Simionescu, Smith, Suzuki, Szymkowiak, Takahashi, Takeo, Tamagawa, Tamura, Tanaka,
  Tanimoto, Tashiro, Terada, Terashima, Tsuboi, Tsujimoto, Tsunemi, Tsuru, T{\"u}mer, Uchida, Uchida, Uchida, Uchiyama, Ueda, Ueda, Uno, Vink, Watanabe, Williams, Yamada, Yamada, Yamaguchi, Yamaoka, Yamasaki, Yamauchi, Yamauchi, Yaqoob, Yoneyama, Yoshida, Yukita, Zhuravleva, Fabian, Nelson, Okabe, Pillepich, Potter, Regamey, Sakai, Shishido, Truong, Wik, \& Zuhone}]{XRISMCollaboration+2025d}
{XRISM Collaboration}, Audard, M., Awaki, H., {et~al.} 2025{\natexlab{a}}, Astrophys. J., 985, L20

\bibitem[{{XRISM Collaboration} {et~al.}(2025{\natexlab{b}}){XRISM Collaboration}, Audard, Awaki, Ballhausen, Bamba, Behar, {Boissay-Malaquin}, Brenneman, Brown, Corrales, Costantini, Cumbee, Diaz~Trigo, Done, Dotani, Ebisawa, Eckart, Eckert, Eguchi, Enoto, Ezoe, Foster, Fujimoto, Fujita, Fukazawa, Fukushima, Furuzawa, Gallo, Garc{\'i}a, Gu, Guainazzi, Hagino, Hamaguchi, Hatsukade, Hayashi, Hayashi, Hell, {Hodges-Kluck}, Hornschemeier, Ichinohe, Ishi, Ishida, Ishikawa, Ishisaki, Kaastra, Kallman, Kara, Katsuda, Kanemaru, Kelley, Kilbourne, Kitamoto, Kobayashi, Kohmura, Kubota, Leutenegger, Loewenstein, Maeda, Markevitch, Matsumoto, Matsushita, McCammon, McNamara, Mernier, Miller, Miller, Mitsuishi, Mizumoto, Mizuno, Mori, Mukai, Murakami, Mushotzky, Nakajima, Nakazawa, Ness, Nobukawa, Nobukawa, Noda, Odaka, Ogawa, Ogorza{\l}ek, Okajima, Ota, Paltani, Petre, Plucinsky, Porter, Pottschmidt, Sato, Sato, Sawada, Seta, Shidatsu, Simionescu, Smith, Suzuki, Szymkowiak, Takahashi, Takeo, Tamagawa, Tamura, Tanaka,
  Tanimoto, Tashiro, Terada, Terashima, Tsuboi, Tsujimoto, Tsunemi, Tsuru, T{\"u}mer, Uchida, Uchida, Uchida, Uchiyama, Ueda, Ueda, Uno, Vink, Watanabe, Williams, Yamada, Yamada, Yamaguchi, Yamaoka, Yamasaki, Yamauchi, Yamauchi, Yaqoob, Yoneyama, Yoshida, Yukita, Zhuravleva, Cui, Ettori, Grayson, Heinrich, McCall, Nelson, Okabe, Omiya, Sarkar, Scannapieco, Sun, Tanaka, Truong, Wik, Zhang, \& Zuhone}]{XrismCollaboration+2025j}
{XRISM Collaboration}, Audard, M., Awaki, H., {et~al.} 2025{\natexlab{b}}, Astrophys. J., 993, L11

\bibitem[{{XRISM Science Team}(2022)}]{XRISMScienceTeam2022}
{XRISM Science Team}. 2022, {{XRISM Quick Reference}}

\bibitem[{Zhuravleva {et~al.}(2019)Zhuravleva, Churazov, Schekochihin, Allen, Vikhlinin, \& Werner}]{Zhuravleva+2019}
Zhuravleva, I., Churazov, E., Schekochihin, A.~A., {et~al.} 2019, Nat. Astron., 3, 832

\bibitem[{ZuHone {et~al.}(2014)ZuHone, Kunz, Markevitch, Stone, \& Biffi}]{ZuHone+2014}
ZuHone, J.~A., Kunz, M.~W., Markevitch, M., Stone, J.~M., \& Biffi, V. 2014, Astrophys. J., 798, 90

\end{thebibliography}

\end{document}